\documentclass[prl,amsmath,amssymb,twocolumn,superscriptaddress]{revtex4-1}

\usepackage{}
\usepackage{amsfonts}

\usepackage[utf8]{inputenc}
\usepackage[T1]{fontenc}
\usepackage[english]{babel}  

\usepackage{times}

\usepackage{amssymb,amsmath,amsfonts,amsthm}
\usepackage{mathtools}
\usepackage{verbatim}
\usepackage{enumerate}
\usepackage{graphicx}
\graphicspath{{figs/}}
\usepackage{hyperref}
\usepackage{bbm}

\usepackage[caption=false]{subfig}

\renewcommand{\baselinestretch}{1}  

\usepackage{color}
\definecolor{dred}{rgb}{.8,0.2,.2}
\definecolor{ddred}{rgb}{.8,0.5,.5}
\definecolor{dblue}{rgb}{.2,0.2,.8}
\definecolor{dgreen}{rgb}{.2,0.5,.2}

\begin{document}

\newcommand{\add}[1]{\textcolor{dred}{*** #1 ***}}
\newcommand{\out}[1]{\textcolor{ddred}{\textbf{[}\emph{#1}\textbf{]}}}
\newcommand{\blue}[1]{\textcolor{dblue}{\textbf{[}#1\textbf{]}}}
\newcommand{\red}[1]{\textcolor{dred}{#1}}
\newcommand{\todo}[1]{\textbf{\underline{\textcolor{dblue}{\textbf{[}#1\textbf{]}}}}}
\newcommand{\tomi}[1]{\textcolor{dgreen}{\textbf{[Tomi: }#1\textbf{]}}}

\theoremstyle{plain}
\newtheorem{theorem}{Theorem}   
\newtheorem{proposition}[theorem]{Proposition}
\newtheorem{lemma}[theorem]{Lemma}
\newtheorem{corollary}[theorem]{Corollary}

\theoremstyle{definition}
\newtheorem{definition}[theorem]{Definition}
\newtheorem{example}[theorem]{Example}
\newtheorem{remark}[theorem]{Remark}

\newcommand{\bra}[1]{\mbox{$\langle #1|$}}
\newcommand{\ket}[1]{\ensuremath{|#1\rangle}}
\newcommand{\braket}[2]{\mbox{$\langle #1|#2\rangle$}}
\newcommand{\ketbra}[2]{\mbox{$|#1\rangle\langle #2|$}}
\newcommand{\iprod}[2]{\ensuremath{\langle #1,#2 \rangle}}

\newcommand{\gate}[1]{\ensuremath{\text{\sc #1}}}
\newcommand{\COPY}[1][]{\ensuremath{\gate{COPY}_{#1}}}

\newcommand{\comm}[2]{\ensuremath{\left[#1, #2\right]}}

\newcommand{\eq}{\Leftrightarrow}

\newcommand{\projector}[1]{\mbox{$|#1\rangle\langle #1|$}}

\newcommand{\x}{\mathbf{x}}
\newcommand{\y}{\mathbf{y}}

\newcommand{\hprod}{\odot}

\newcommand{\I}{\openone}     
\newcommand{\R}{{\mathbb R}}  
\newcommand{\hilb}[1]{\ensuremath{\mathcal{#1}}} 
\newcommand{\swap}{{\sf{SWAP}}}
\newcommand{\ie}{i.e.}

\newcommand{\hs}{\kappa}

\newcommand{\be}{\begin{equation}}
\newcommand{\ee}{\end{equation}}



\def\1#1{{\bf #1}}
\def\2#1{{\cal #1}}
\def\7#1{{\mathbb #1}}

\newcommand{\bea}{\begin{eqnarray}}
\newcommand{\eea}{\end{eqnarray}}

\newcommand{\half}{\mbox{$\textstyle \frac{1}{2}$}}
\newcommand{\quarter}{\mbox{$\textstyle \frac{1}{4}$}}

\newcommand{\brakets}[2]{\langle\, #1\,|\,#2\,\rangle}
\newcommand{\bracket}[3]{\left\langle #1 \left| #2 \right| #3 \right\rangle}
\newcommand{\brackets}[3]{\langle #1 | #2 | #3 \rangle}
\newcommand{\proj}[1]{\ket{#1}\bra{#1}}
\newcommand{\av}[1]{\langle #1\rangle}
\newcommand{\outprod}[2]{\ket{#1}\bra{#2}}
\newcommand{\op}[2]{|#1\rangle \langle #2|}
\newcommand{\tr}{\textrm{tr}}
\newcommand{\od}[2]{\frac{\mathrm{d} #1}{\mathrm{d} #2}}
\newcommand{\pd}[2]{\frac{\partial #1}{\partial #2}}
\newcommand{\dt}[1]{\frac{\partial #1}{\partial t}}

\newcommand{\an}[1]{\hat{#1}}
\newcommand{\cre}[1]{\hat{#1}^\dag}
\newcommand{\vac}{\ket{\textrm{vac}}}

\newcommand{\cc}{\textrm{c.c.}}

\newcommand{\identity}{\mathbbm{1}}

\newcommand{\up}{\uparrow}
\newcommand{\down}{\downarrow}

\renewcommand{\Re}{\mathfrak{Re}}
\renewcommand{\Im}{\mathfrak{Im}}

\renewcommand{\baselinestretch}{1}


\def\thesection{%
\Roman{section}}%
\def\thesubsection{%
\Alph{subsection}}%
\def\thesubsubsection{%
\arabic{subsubsection}}%
\def\theparagraph{%
\Roman{paragraph}}%
\def\thesubparagraph{%
\theparagraph.\Roman{subparagraph}}%
\setcounter{secnumdepth}{5}%

\title{Quantum State and Process Tomography via Adaptive Measurements}
\author{Hengyan Wang}
\affiliation{Hefei National Laboratory for Physical Sciences at Microscale and Department of Modern Physics, University of Science
and Technology of China, Hefei, Anhui 230026, China}

\author{Wenqiang Zheng}
\affiliation{Department of Applied Physics, Zhejiang University of Technology, Hangzhou, Zhejiang 310023, China}

\author{Nengkun Yu}
\email{nengkunyu@gmail.com}
\affiliation{Institute for Quantum Computing,
University of Waterloo, Waterloo N2L 3G1, Ontario, Canada}
\affiliation{Department of Mathematics \& Statistics, University of
  Guelph, Guelph N1G 2W1, Ontario, Canada}
\affiliation{Centre for Quantum Computation \& Intelligent Systems,
   Faculty of Engineering and Information Technology, University of
   Technology Sydney, NSW 2007, Australia}

\author{Keren Li}
\affiliation{Institute for Quantum Computing,
University of Waterloo, Waterloo N2L 3G1, Ontario, Canada}
\affiliation{State Key Laboratory of Low-Dimensional Quantum Physics and Department of Physics, Tsinghua University, Beijing 100084, China}

\author{Dawei Lu}
\email{d29lu@uwaterloo.ca}
\affiliation{Institute for Quantum Computing,
University of Waterloo, Waterloo N2L 3G1, Ontario, Canada}

\author{Tao Xin}
\affiliation{Institute for Quantum Computing,
University of Waterloo, Waterloo N2L 3G1, Ontario, Canada}
\affiliation{State Key Laboratory of Low-Dimensional Quantum Physics and Department of Physics, Tsinghua University, Beijing 100084, China}

\author{Carson Li}

\affiliation{Institute for Quantum Computing,
University of Waterloo, Waterloo N2L 3G1, Ontario, Canada}
\affiliation{Department of Physics, University of
  Guelph, Guelph, Ontario, Canada}

\author{Zhengfeng Ji}
\affiliation{Centre for Quantum Computation \& Intelligent Systems,
   Faculty of Engineering and Information Technology, University of
   Technology Sydney, NSW 2007, Australia}
\affiliation{State Key Laboratory of Computer Science, Institute of
  Software, Chinese Academy of Sciences, Beijing, China}%

\author{David Kribs}
\affiliation{Department of Mathematics \& Statistics, University of
  Guelph, Guelph N1G 2W1, Ontario, Canada}
\affiliation{Institute for Quantum Computing,
University of Waterloo, Waterloo N2L 3G1, Ontario, Canada}

\author{Bei Zeng}
\email{zengb@uoguelph.ca}
\affiliation{Institute for Quantum Computing,
University of Waterloo, Waterloo N2L 3G1, Ontario, Canada}
\affiliation{Department of Mathematics \& Statistics, University of
  Guelph, Guelph N1G 2W1, Ontario, Canada}
\affiliation{Canadian Institute for Advanced Research, Toronto,
  Ontario, Canada}

\author{Xinhua Peng}
\email{xhpeng@ustc.edu.cn}
\affiliation{Hefei National Laboratory for Physical Sciences at Microscale and Department of Modern Physics, University of Science
and Technology of China, Hefei, Anhui 230026, China}

\author{Jiangfeng Du}
\affiliation{Hefei National Laboratory for Physical Sciences at Microscale and Department of Modern Physics, University of Science
and Technology of China, Hefei, Anhui 230026, China}

\begin{abstract}
We investigate quantum state tomography (QST) for pure states and quantum process tomography (QPT)
for unitary channels via \textit{adaptive} measurements. For a quantum
system with a $d$-dimensional Hilbert space, we first propose an adaptive protocol where only $2d-1$ measurement outcomes are used to accomplish the QST for \textit{all} pure states. This idea is then extended to study QPT for unitary channels, where an adaptive unitary process tomography (AUPT) protocol of $d^2+d-1$ measurement outcomes is constructed for any unitary channel. We experimentally implement the AUPT protocol in a 2-qubit nuclear magnetic resonance system. We examine the performance of the AUPT protocol when applied to Hadamard gate, $T$ gate ($\pi/8$ phase gate), and controlled-NOT gate, respectively, as these gates form the universal gate set for quantum information processing purpose. As a comparison, standard QPT is also implemented for each gate. Our experimental results show that the AUPT protocol that reconstructing unitary channels via adaptive measurements significantly reduce the number of experiments required by standard QPT without considerable loss of fidelity.
\end{abstract}

\maketitle

\section{Introduction}
The problem of how many measurements are needed to determine a wave function of a quantum system is a nontrivial task even in principle, and has attracted considerable attention over the history of the subject. Originally raised by Pauli in 1933, the problem was framed as whether the probability distribution of position and momentum is enough to determine the wave function~\cite{pauli1933allgemeinen}. Subsequently, various versions of the problem and many different approaches have been explored~\cite{weigert1992pauli,amiet1999reconstructing}.

For a system of finite dimension $d$ with Hilbert space $\mathcal{H}_d$, a normalized pure state $\ket{\psi_d}\in\mathcal{H}_d$ is specified by $2d-2$ real parameters. To measure any observable $\mathbf{A}$ on the ensemble of identical copies of the states $\ket{\psi_d}\in\mathcal{H}_d$, the expectation $\bra{\psi}\mathbf{A}\ket{\psi}$ is returned. In order to determine an arbitrary $\ket{\psi}$, at least $2d-2$ such observables need to be measured.

The development of quantum information science has shed new light on the problem~\cite{flammia2005minimal,finkelstein2004pure,GLF+10,CPF+10,heinosaari2013quantum,chen2013uniqueness,baldwin2015informational,li2015fisher,ma2016pure,xin2016quantum,carmeli2016efficient}, which can be rephrased by quantum state tomography (QST) for pure states in the language of quantum information. In particular, the precise meaning of the word `determine' is clarified, where two important scenarios are considered~\cite{chen2013uniqueness}.
The first scenario is whether the measurement results uniquely determine the pure state among all pure states (UDP, i.e. no other pure states can give the same measurement result) or among all states (UDA, i.e. no other states, pure or mixed, can give the same measurement result). The latter is an arguably stronger requirement and gaps are found between the number of measurements needed for UDP and UDA.
The second scenario is whether the measurement results determine (UDP or UDA) all pure states (i.e. any state can be reconstructed unambiguously) or just generic pure states (i.e. almost all pure states are determined except a set of states that are of measure zero). The former is an arguably stronger requirement and gaps are found between the number of measurements needed for all pure states and generic pure states~\cite{chen2013uniqueness}. In columns $2$ and $3$ of Table~\ref{tb1}, we summarize the best known number of measurements needed
for UDP/UDA for all states in the row of starting with ``All'' and for generic states in the row of starting with ``Generic''.

Now we naturally extend the above problem of QST for pure states to quantum process tomography (QPT) for unitary channels. QPT for unitary channels has the goal of determining an unknown unitary operation. A $d\times d$ unitary operation has $d^2-1$ real parameters, compared to a general quantum channel on a $d$-dimensional system that have $d^4-d^2$ real parameters. In~\cite{gutoski2014process}, it is shown that $4d^2-2d-4$ measurements are sufficient to identify a unitary channel among $all$ unitary channels, non-adaptively. Their method is based on the state tomography of the corresponding Choi matrix of the unitary channel.
Ref.~\cite{Baldwin2014} provides a nonadaptive method of unitary tomography using $d^2+d-1$ measurements, whereas it does not works for all unitary channels but for almost all unitary channels (i.e. works for `generic channels'). Their method is based on the fact that each column of the unitary matrix $U$ can be determined by QST for an input state that is a computational basis state, and the relative phases between any two columns of $U$ can be further determined by QST for some input states that are superpositions of computational basis states. We summarize these results in column $4$ of Table~\ref{tb1}.

All the above-mentioned protocols for either QST or QPT are non-adaptive, that is, the observables to be measured  are fixed once chosen. One can also consider \textit{adaptive} measurement by allowing measurements that are determined by the results of the previous measurements. There has been such trials along this direction, and a $5d$ measurements protocol via adaptive measurements are discussed in~\cite{goyeneche2014five}, for UDA all pure states. One important open question is what are the minimum number of measurements needed for QST of \textit{all} pure states, and for QPT of \textit{all} unitary channels.

In this work, we study QST for pure states and QPT for unitary channels, using adaptive measurements. For QST, we show that $2d-1$ measurements are enough to UDA (hence UDP)
\textit{all} pure states, by adaptive measurements. This is a significant improvement over the $4d-5$ lower bound for UDP using non-adaptive measurements \cite{heinosaari2013quantum}.
We then further apply our protocol to study QPT of unitary channels, and show that $d^2+d-1$ measurements are sufficient to reconstruct \textit{all} unitary channels when adaptive scheme is allowed.

\begin{center}
\begin{table}[h!]
\label{tb1}
\begin{tabular}{ | c | c| c| c|}
\hline
& UDP & UDA & UPT \\
\hline
All (Nonadaptive) & $4d-5$~{\cite{heinosaari2013quantum}} & $5d-7$~\cite{chen2013uniqueness} & $4d^2-2d-4$~{\cite{gutoski2014process}} \\
\hline
Generic (Nonadaptive)& $2d-1$~\cite{finkelstein2004pure} & $2d-1$~\cite{baldwin2015informational} & $d^2+d-1$~{\cite{Baldwin2014}}\\
\hline
All (Adaptive) & $2d-1$ & $2d-1$ &  $d^2+d-1$\\
\hline
\end{tabular}
\caption{Columns $2$ and $3$: A summary of the best known number of measurements needed
for UDP/UDA all states in the row of starting with ``All'' and for generic states in the row of starting with ``Generic',
by nonadaptive measurements.
The number of measurements needed for UDP/UDA all states by adaptive measurements, based on the results obtained in this work, is in the last row starting with ``All (Adaptive)''. Column $4$: A summary of the best known number of measurements for unitary process tomography (UPT) for all unitary channels in the row starting with ``All'' and for generic unitary channels in the row starting with ``Generic', for nonadaptive measurements. The number of measurements to determine all unitary channels by adaptive measurements, based on the results in this work, is in the last row starting with ``All (Adaptive)''.}
\end{table}
\end{center}

We organize our paper as follows: in Sec.~\ref{sec:theory}, we discuss an adaptive protocol that UDA (hence UDP) for \textit{all} pure states with measuring $2d-1$ observables; we then apply this protocol on QPT of unitary channels by measuring $d^2+d-1$ observables. In Sec. ~\ref{sec:ExpPro}, we discuss an adaptive experimental protocol of QPT for two-qubit unitary channels. In Sec.~\ref{sec:Exp}, we implement the experimental protocol in a two-qubit NMR system. Our experimental results are discussed in Sec.~\ref{sec:results}, followed by a brief conclusion in Sec.~\ref{sec:conclusion}.

\section{Adaptive protocols for quantum state and process tomography}
\label{sec:theory}

In this section, we discuss adaptive protocols for QST and QPT. We start from the case of QST for pure states in Sec.~\ref{sec:AST}, then further extend it to QPT for unitary channels in Sec.~\ref{sec:AU}.

\subsection{Adaptive Pure State Tomography}
\label{sec:AST}
In this subsection, we propose an adaptive pure state tomography (APST) protocol for $d$-dimensional pure states using at most $2d-1$ observables.

The state space we considered is spanned by orthogonal basis $\{\ket{i}:0\leq i\leq d-1\}$.
Suppose the quantum state is
\begin{equation}\label{state}
\ket{\psi}=\sum_{n=0}^{d-1}\alpha_n\ket{n}.
\end{equation}
The goal of tomography is to obtain all $\alpha_n$'s for $0\leq n\leq d-1$, and the APST protocol is given as follows:

\textbf{Step 1}. Measure $\ket{\psi}$ using measurements $E_0,E_1,\cdots$ sequentially until $\tr(\ket{\psi}\bra{\psi} E_k)$ is non-zero, where $E_k=\op{k}{k}$. The goal is to find the smallest $k$ such that $\alpha_k\neq 0$. Hence this step costs $k+1$ measurements, and the state becomes
\begin{equation}\label{state}
\ket{\psi}=\sum_{n=k}^{d-1}\alpha_n\ket{n},
\end{equation}
where the summation starts from $n=k$ now. Without loss of generality, we assume that $\alpha_k=\sqrt{\tr(\ket{\psi}\bra{\psi} E_k)}$ is real since the global phase of a quantum state is ignorable.

\textbf{Step 2}. Measure $\ket{\psi}$ using measurements $F_n,G_n,\cdots$ for all $k<n<d$ with Hermitian $F_n+G_n=\op{n}{k}+\op{k}{n}$ and $F_n-G_n=i(\op{n}{k}-\op{k}{n})$. The goal of this step is to obtain $\alpha_n$ for all $n\geq k$ by employing the coherence between $\ket{k}$ and $\ket{n}$. This step costs $2(d-k-1)$ measurements.

In total, the number of measurements is $2d-k-1$ which is no more than $2d-1$, depending on when we have measured the non-zero $\alpha_k$ for the first time. In terms of density matrix, our protocol actually provides the ($k+1$)-th row of $\op{\psi}{\psi}$.

In the following, we analyze this protocol and show that it indeed accomplishes the task of QST for pure states. In other words, one can compute each $\alpha_n$ according to the outcomes of this protocol.

We first show that it is UDP.
After step 1, we know that $\ket{\psi}=\sum_{n=k}^{d-1}\alpha_n\ket{n}$.
After step 2, we have
\bea
\langle\psi|(F_n+G_n)\ket{\psi}&=&\alpha_k\alpha_n+\alpha_k\bar{\alpha_n}, \\ \nonumber
\langle\psi|(F_n-G_n)\ket{\psi}&=&i(\alpha_k\alpha_n-\alpha_k\bar{\alpha_n}).
 \eea
As we have assumed that $\alpha_k$ is real in step 1, it is obvious that $\bar{\alpha_k}\alpha_n=\alpha_k\alpha_n$ for all $n>k$. Therefore, we can calculate the exact value of $\alpha_n$ since we know the non-zero $\alpha_k$ and $\alpha_k\alpha_n$ from our measurements. It means we have the complete information of $\ket{\psi}$ if we know it is pure.

Next we prove that this APST protocol is not only UDP, but also UDA.
To see this, we need to show that if another quantum state $\rho$ which gives the same results as $\ket{\psi}$, $\rho$ can only be $\op{\psi}{\psi}$.

Assume there exists another quantum state $\rho$ that has the same measurement results compared to $\ket{\psi}$. So for $n<k$, we have
\begin{align}\label{apst}
\tr(\rho\op{n}{n})&=\tr(\op{\psi}{\psi}\op{n}{n})=0,\\
\tr(\rho\op{k}{k})&=\tr(\op{\psi}{\psi}\op{k}{k})=\alpha_k^2.
\end{align}
For $k\leq n\leq d-1$, we have
\begin{align}
\tr(\rho\op{k}{n})=\tr(\op{\psi}{\psi}\op{k}{n})=\alpha_k\alpha_n.
\end{align}
In other words, our protocol actually outputs the first non-zero row of $\rho$, which is the $(k+1)$-th row. This row of $\rho$ equals to the $(k+1)$-th row of $\op{\psi}{\psi}$.

As $\rho$ is semi-definite positive, we can suppose $$\rho=\sum_{j=0}^{d-1} \op{\phi_j}{\phi_j}$$ with unnormalized $\ket{\phi_j}=(\beta_{0,j},\cdots,\beta_{d-1,j})^T$.
According to Eq. (\ref{apst}) and the semi-definite positive property of $\rho$, we know that the first $k$ rows of $\rho$ are all zero, namely, $\beta_{r,j}=0$ for all $r<k$.

Without loss of generality, we can assume that $\beta_{k,0}\neq 0$ and $\beta_{k,j}=0$ for all $j>0$. This property helps us to show $\op{\psi}{\psi}=\op{\phi_0}{\phi_0}$.
To achieve such a decomposition, we first observe that
$$\op{\varphi_1}{\varphi_1}+\op{\varphi_2}{\varphi_2}=\op{\varsigma_1}{\varsigma_1}+\op{\varsigma_2}{\varsigma_2}$$
where
\begin{align*}
\ket{\varsigma_1}=u\ket{\varphi_1}+v\ket{\varphi_2},\\
\ket{\varsigma_2}=\bar{v}\ket{\varphi_1}-\bar{u}\ket{\varphi_2},
\end{align*}
with $|u|^2+|v|^2=1$.

Apply this on $\op{\phi_0}{\phi_0}+\op{\phi_1}{\phi_1}$ by choosing $u,v$ appropriately, we can always achieve $\beta_{k,1}=0$. Employing this argument recursively on $\op{\phi_0}{\phi_0}+\op{\phi_j}{\phi_j}$, we can similarly have $\beta_{k,j}=0$ for all $j>0$. Then, the $(k+1)$-th row of $\sum_{j=1}^{d-1} \op{\phi_j}{\phi_j}$ are all zero.

According to $\rho=\sum_{j=0}^{d-1} \op{\phi_j}{\phi_j}$, we observe that the $(k+1)$-th row of $\rho$ equals to the $(k+1)$-th row of $\op{\phi_0}{\phi_0}$. Thus, the $(k+1)$-th row of $\op{\phi_0}{\phi_0}$ equals to the $(k+1)$-th row of $\op{\psi}{\psi}$.
Therefore, $\ket{\phi_0}$ equals to $\ket{\psi}$ up to a global phase, which means $\op{\phi_0}{\phi_0}=\op{\psi}{\psi}$. Thus, $$\tr(\sigma)=\tr(\rho)-\tr(\op{\psi_0}{\psi_0})=0$$ where $$\sigma=\rho-\op{\psi_0}{\psi_0}=\sum_{i=1}^{d-1} \op{\phi_i}{\phi_i}.$$ That is $\sigma=0$, and $$\rho=\op{\psi_0}{\psi_0}=\op{\psi}{\psi}.$$

This verifies our claim that our APST protocol is UDA and uses only $2d-1$ measurements.


\subsection{Adaptive Unitary Process Tomography}
\label{sec:AU}

In this subsection, the idea of APST is generalized to deal with the adaptive unitary process tomography (AUPT). We notice that the unitary map $U$ can be written as a transformation from the orthonormal basis $\{\ket{n}\}$ to its image basis $\{\ket{u_n}\}$,
\begin{equation}\label{unitary}
U=\sum_{n=0}^{d-1}\ket{u_n}\bra{n}.
\end{equation}
The task of QPT for a unitary map is to fully characterize the basis $\{\ket{u_n}\}$ and the relative phases $\{\ket{u_n}\bra{n}\}$, and our AUPT protocol consists of $d$ steps as follows,

\textbf{Step $1$}. Implement QST for $\ket{u_0}=U\ket{0}$. We use the APST protocol in the previous subsection to characterize $\op{u_0}{u_0}$. This step costs at most $2d-1$ measurements.

\textbf{Step $2$}. Implement QST for $$U\ket{+}=U(\ket{0}+\ket{1})/\sqrt{2}=(\ket{u_0}+\ket{u_1})/\sqrt{2}.$$ The goal of this step is to tomography $\ket{u_1}$ and to obtain the relative phase between $\ket{u_0}$ and $\ket{u_1}$ simultaneously. This can be done by obtaining $$U\op{+}{+}U^{\dag}=\op{\varphi}{\varphi}$$ using our APST protocol, so that we can construct $\ket{u_1}$. To see this, we notice that $(\ket{u_0}+\ket{u_1})/\sqrt{2}=e^{i\gamma}\ket{\varphi}$. Observe that
the inner product of $(\ket{u_0}+\ket{u_1})/\sqrt{2}$ and $\ket{u_0}$ is $1/\sqrt{2}$. This indicates that the phase information of $\gamma$ is obtained. Then, the information of $\ket{u_1}$ is obtained.

Moreover, we observe that
$$\op{u_0+u_1}{u_0+u_1}=\op{u_0+e^{i\theta}u_1}{u_0+e^{i\theta}u_1}$$
has the only solution that $e^{i\theta}=1$. That implies that the information of the relative phase between $\ket{u_0}$ and $\ket{u_1}$ is obtained completely.
We choose a basis $$\{\ket{v_{0,n}}\bra{n}+\ket{n}\bra{v_{0,n}},i(\ket{v_{0,n}}\bra{n}-\ket{n}\bra{v_{0,n}}), 0\leq n\leq d-1\}$$ with $\ket{v_{0,0}}=\ket{u_0}$. and then apply this basis using step 2 in the APST protocol to obtain $\ket{u_1}$.

This step costs $2d-2$ measurements since we already know the amplitude of $\ket{u_0}$ is $1/\sqrt{2}$.

\textbf{Step $j$}. Implement QST for $$U(\ket{0}+\ket{j-1})/\sqrt{2}=(\ket{u_0}+\ket{u_{j-1}})/\sqrt{2}.$$
The goal of this step is to tomography $\ket{u_{j-1}}$ and obtain the relative phase between $\ket{u_0}$ and $\ket{u_{j-1}}$ simultaneously.
The procedure to obtain $\ket{u_{j-1}}$ is similar to step 2 by choosing a basis $$\{\ket{v_{1,n}}\bra{n}+\ket{n}\bra{v_{1,n}},i(\ket{v_{1,n}}\bra{n}-\ket{n}\bra{v_{1,n}}), 0\leq n\leq d-1\}$$ with $\ket{v_{1,r}}=\ket{u_r}$ for all $r\leq j-2$, and applying it on the APST protocol.

This step costs $2(d-j+1)$ measurements, since we already know the amplitude of $\ket{u_0}$ is $1/\sqrt{2}$ for state $(\ket{u_0}+\ket{u_{j-1}})/\sqrt{2}$ , and the amplitudes of $\ket{u_1},\cdots,\ket{u_{j-2}}$ are all zero.

The above steps keep going until step $d$, in which two measurements are required and the complete information of $U$ is obtained from the outcomes of the $d$ steps. This AUPT protocol thus uses
$2d-1+\sum_{j=2}^{d} 2(d-j+1)=d^2+d-1$ measurements.

\section{Experimental protocol}
\label{sec:ExpPro}

In this section, we show how to apply our AUPT protocol to characterize unitary channels  (as discussed in Sec.~\ref{sec:AU}) in a 2-qubit NMR system, and its complexity, i.e. the number of measurements in terms of Pauli operators. As a comparison, we also briefly review how to implement a standard QPT and the complexity. The extension of our protocol to arbitrary sizes is straightforward.

\subsection{Standard QPT}

First let us recall the procedure of a 2-qubit standard QPT. Suppose $\mathcal{U}$ is the $2$-qubit unitary gate that we want to implement in practice.
Due to the inevitable experiment errors, the real quantum channel in the laboratory is no longer unitary, but still some completely positive trace-preserving (CPTP) operation, denoted by $\Lambda$. In NMR and most of ensemble systems, it is convenient to prepare and measure Pauli observables, hence we use the representation of Pauli observables to describe such a 2-qubit channel $\Lambda$. Note that this description is equivalent to the Choi matrix representation and they can be easily transformed to each other \cite{chow2012universal}.

Therefore, $\Lambda$ can be written in the way of mapping Pauli group to Pauli group so that
\begin{equation}
\Lambda \left( {\begin{array}{*{20}{c}}
{XX}\\
{XY}\\
{...}\\
{II}
\end{array}} \right) = \left( {\begin{array}{*{20}{c}}
{{p^1_1}}&{{p^2_1}}&{...}&{{p^{15}_1}}&{{p^{16}_1}}\\
{{p^1_2}}&{{p^2_2}}&{...}&{{p^{15}_2}}&{{p^{16}_2}}\\
{...}&{...}&{...}&{{...}}&{{...}}\\
{{p^1_{16}}}&{{p^2_{16}}}&{...}&{{p^{15}_{16}}}&{{p^{16}_{16}}}
\end{array}} \right)\left( {\begin{array}{*{20}{c}}
{XX}\\
{XY}\\
{...}\\
{II}
\end{array}} \right),
\label{StdQPT}
\end{equation}
where all elements $p^i_j$ ($1\leq i, j \leq 16$) are real. To reconstruct $\Lambda$ in NMR, we firstly prepare the initial state as $XX$, and
then apply $\Lambda$ on it. The output state is thus $p^1_1XX+p^1_2XY+...+p^1_{16}II$. By doing a full state tomography in 15 experiments, i.e. measuring each $p^1_j$ ($1\leq j \leq 15$, since $p^1_{16}$ can only be computed via the normalization condition), we can obtain the first column of $\Lambda$. To fully characterize $\Lambda$, the above procedure needs to be repeated by $16$ times, with each time preparing a distinct Pauli input state out of $\{XX,XY,...,II\}$. So the total number of experiments to reconstruct a 2-qubit channel $\Lambda$ is $16\times15=240$.

\subsection{AUPT}
If we assume $\mathcal{U}$ is still unitary when applied in practice, the total number of experiments can be reduced significantly.
Due to the experiment errors, let us denote $\mathcal{V}$ as the real channel, which is still unitary but deviates from the desired $\mathcal{U}$. As unitary operators do not change the purities when applied on quantum states, it is convenient to consider the map from pure states to pure states. Explicitly, the map of $\mathcal{V}$ can be written as
\begin{equation}
\mathcal{V}\left( {\begin{array}{*{20}{c}}
{\left| {00} \right\rangle }\\
{\left| {01} \right\rangle }\\
{\left| {10} \right\rangle }\\
{\left| {11} \right\rangle }
\end{array}} \right) = \left( {\begin{array}{*{20}{c}}
{{\alpha _1}}&{{\beta _1}}&{{\gamma _1}}&{{\delta _1}}\\
{{\alpha _2}}&{{\beta _2}}&{{\gamma _2}}&{{\delta _2}}\\
{{\alpha _3}}&{{\beta _3}}&{{\gamma _3}}&{{\delta _3}}\\
{{\alpha _4}}&{{\beta _4}}&{{\gamma _4}}&{{\delta _4}}
\end{array}} \right)\left( {\begin{array}{*{20}{c}}
{\left| {00} \right\rangle }\\
{\left| {01} \right\rangle }\\
{\left| {10} \right\rangle }\\
{\left| {11} \right\rangle }
\end{array}} \right),
\label{UnitaryQPT}
\end{equation}
where the elements in $\mathcal{V}$ are all complex numbers. Similarly to standard QPT, in experiment we firstly prepare $\ket{00}$ and then apply $\mathcal{V}$. The output quantum state is still pure since
\begin{equation}
\mathcal{V}\left| {00} \right\rangle  = {\alpha _1}\left| {00} \right\rangle  + {\alpha _2}\left| {01} \right\rangle  + {\alpha _3}\left| {10} \right\rangle  + {\alpha _4}\left| {11} \right\rangle .
\label{V00}
\end{equation}
Now the problem of characterizing a unitary channel converts to the QST of a pure state.
First, we can use three measurements of the diagonal elements combined with the normalization condition to get $\left| {{\alpha _1}} \right|, \left| {{\alpha _2}} \right|, \left| {{\alpha _3}} \right|$ and $\left| {{\alpha _4}} \right|$. Then we need to measure the relative phase between all the $\alpha$'s.  Specifically,
we pick out the maximal $\left| {{\alpha _i}} \right|$ and set its phase as zero. Without loss of generality, assume $\left| {{\alpha _1}} \right|$ is the largest one and set it as reference. To measure, for instance, the relative phase $\theta_{\alpha_2}$ between $\alpha_1$ and $\alpha_2$, is equivalent to extracting the phase between $\ket{00}$ and $\ket{01}$ in experiment, which requires two measurements of $X$ and $Y$ on the second qubit. Analogously, the relative phase $\theta_{\alpha_3}$ and $\theta_{\alpha_4}$ can be measured with four more experiments. Therefore, the total number of experiments to extract the values of $\alpha$ in the first column is nine, with three for moduli and six for relative phases. As $\mathcal{V}$ contains four columns, this procedure is repeated by four times that necessitates 36 experiments, by preparing the input state as $\left|00\right\rangle$, $\left|01\right\rangle$, $\left|10\right\rangle$ and $\left|11\right\rangle$, respectively.

However, the above procedure cannot provide the information of the relative phases between columns, as we have set the phase of the maximal element in each column as zero, but quantum mechanics merely allows one ignorable global phase. So the next step is to determine these relative phases between columns. Without loss of generality, assume $\alpha_1$ is real. To measure the relative phase $\theta_{\alpha\beta}$ between the $\alpha$ column and $\beta$ column, one can adopt the idea of interferometers. Explicitly, prepare the superposition ${{\left( {\left| {00} \right\rangle  + \left| {01} \right\rangle } \right)} \mathord{\left/
 {\vphantom {{\left( {\left| {00} \right\rangle  + \left| {01} \right\rangle } \right)} {\sqrt 2 }}} \right.
 \kern-\nulldelimiterspace} {\sqrt 2 }}$ as the input state and apply $\mathcal{V}$, so that
\begin{equation}
\begin{split}
&\mathcal{V}\left( {\left| {00} \right\rangle  + \left| {01} \right\rangle } \right)/{\sqrt 2 }  \\
 = &\left( {{\alpha _1}\left| {00} \right\rangle  + {\alpha _2}\left| {01} \right\rangle  + {\alpha _3}\left| {10} \right\rangle  + {\alpha _4}\left| {11} \right\rangle } \right)/{\sqrt 2 } + \\
&{e^{i{\theta _{\alpha \beta }}}}\left( {{\beta _1}\left| {00} \right\rangle  + {\beta _2}\left| {01} \right\rangle  + {\beta _3}\left| {10} \right\rangle  + {\beta _4}\left| {11} \right\rangle } \right)/{\sqrt 2 } \\
 = &[\left( {{\alpha _1} + {e^{i{\theta _{\alpha \beta }}}}{\beta _1}} \right)\left| {00} \right\rangle  + \left( {{\alpha _2} + {e^{i{\theta _{\alpha \beta }}}}{\beta _2}} \right)\left| {01} \right\rangle\\
 &+ \left( {{\alpha _3} + {e^{i{\theta _{\alpha \beta }}}}{\beta _3}} \right)\left| {10} \right\rangle  + \left( {{\alpha _4} + {e^{i{\theta _{\alpha \beta }}}}{\beta _4}} \right)\left| {11} \right\rangle]/{\sqrt 2 }.
\end{split}
\label{U0001_1}
\end{equation}
In practice, we can measure the relative phase $\theta_{exp}$ between $\left|00\right\rangle$ and $\left|01\right\rangle$ via two experiments, and the desired $\theta_{\alpha\beta}$ can be obtained by solving the following equation
\begin{equation}
\text{phase}\left( {{\alpha _1} + {e^{i{\theta _{\alpha \beta }}}}{\beta _1},{\alpha _2} + {e^{i{\theta _{\alpha \beta }}}}{\beta _2}} \right) = {\theta _{exp }},
\label{phasemeasure}
\end{equation}
where $\text{phase}(A,B)$ means the relative phase between two complex numbers $A$ and $B$, and all $\alpha$ and $\beta$ values have been obtained in the last step.
Similarly, the relative phases $\theta_{\alpha\gamma}$ and $\theta_{\alpha\delta}$ can be obtained through preparing  $(\left|00\right\rangle+\left|10\right\rangle)/\sqrt{2}$ and $(\left|00\right\rangle+\left|11\right\rangle)/\sqrt{2}$, applying $\mathcal{V}$, and measuring the corresponding phases. This step thus consists of six experiments to acquire three relative phases between columns in $\mathcal{V}$.

In total, we need $36+6=42$ experiments to characterize a 2-qubit unitary process $\mathcal{V}$ via the AUPT protocol, significantly less than the standard QPT which requires 210 experiments.

\section{Experimental implementation in NMR}
\label{sec:Exp}

\begin{figure}[htb]%
\begin{center}
\includegraphics[width= .9\columnwidth]{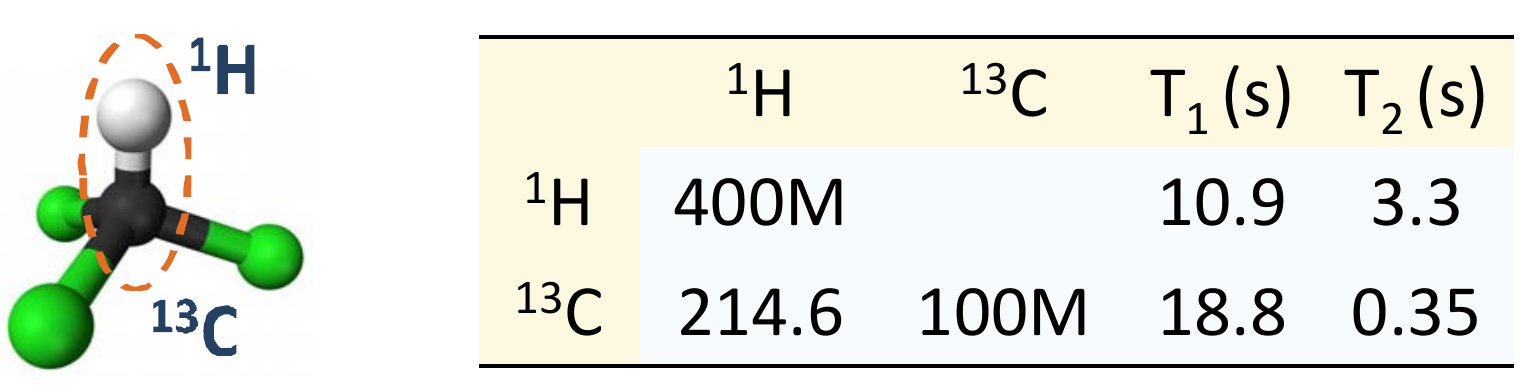}%
\end{center}
\setlength{\abovecaptionskip}{-0.1cm}
\caption{Molecular structure of the 2-qubit sample $^{13}$ C-labeled Chloroform. ${}^1$H and ${}^{13}$C are encoded as qubit $1$ and qubit $2$, respectively. The table on the right summarizes the Hamiltonian parameters at room temperature, including the Larmor frequencies (diagonal, in hertz), the J-coupling strength (off-diagonal, in hertz) and the relaxation time scales $T_1$ and $T_2$. }\label{fig:molecule}
\end{figure}

\begin{figure*}[htb]%
\begin{center}
\includegraphics[width= 2\columnwidth]{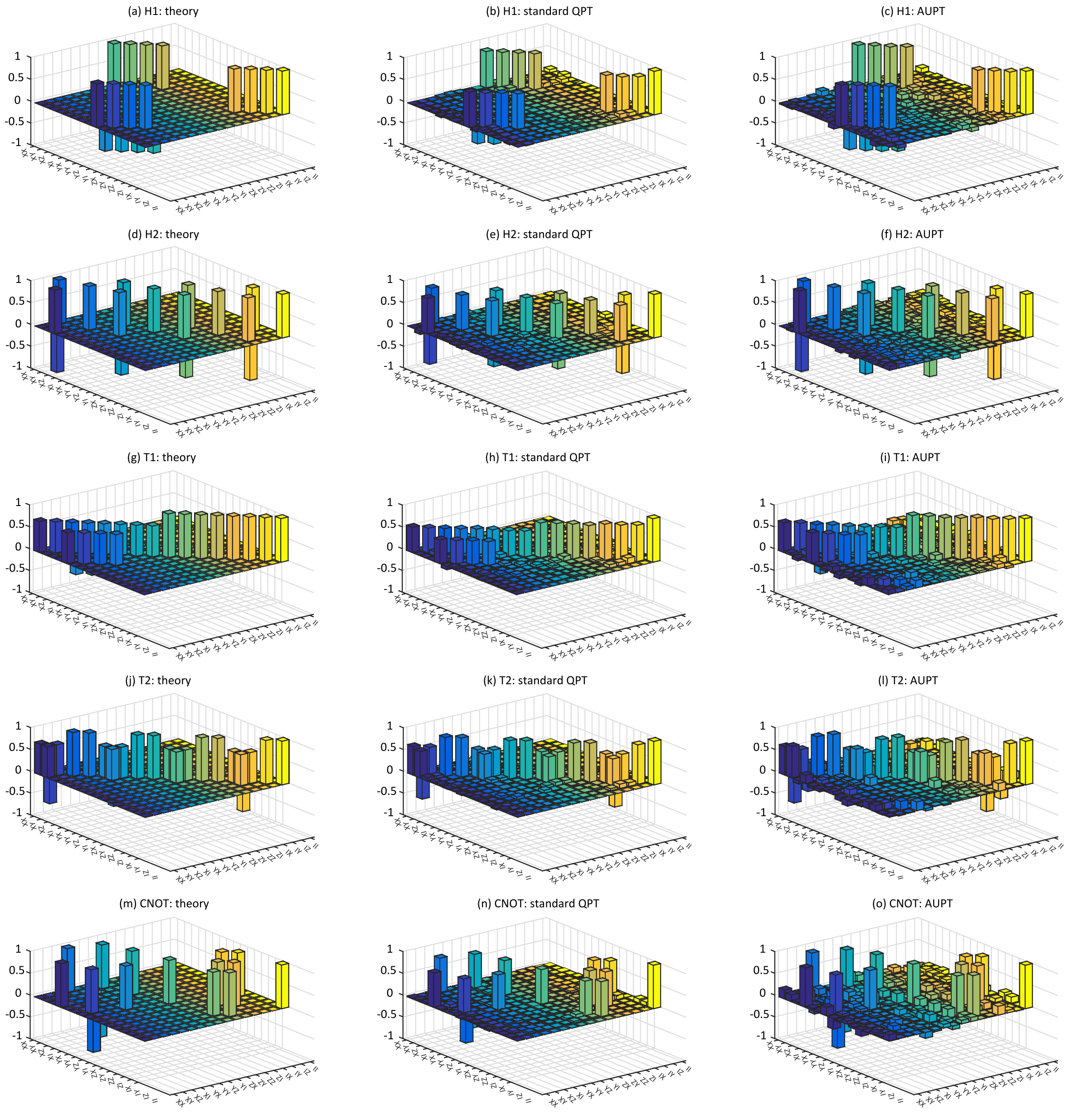}%
\end{center}
\setlength{\abovecaptionskip}{-0.1cm}
\caption{Experimental results of the five gates $H_1$, $H_2$, $T_1$, $T_2$ and $\rm{CNOT}_{12}$ via the standard QPT (middle column) and AUPT protocols (right column), as well as the theoretical results (left column). The five rows correspond to the five gates, respectively. In each subfigure, the $y$-axis describes the input state in Eq. (\ref{StdQPT}), and $x$-axis describes the output state in the Pauli basis after applying the current channel. $z$-axis shows the values of the coefficients of the output state in Pauli basis.}\label{fig:results}
\end{figure*}

Now we turn to the experimental demonstration of the AUPT protocol for 2-qubit unitary gates in the NMR system.
Five elementary gates $H_1=H\otimes I$, $H_2=I\otimes H$, $T_1=T\otimes I$, $T_2=I\otimes T$ and $\mathrm{CNOT}_{12}$ are chosen due to the fact that any 2-qubit quantum circuit can be decomposed into these five gates in arbitrary accuracy \cite{nielsen2010quantum}. $I$ is the identity operator, and the Hadamard gate $H$, $\pi/8$ gate $T$ and controlled-not gate $\mathrm{CNOT}_{12}$ are
\begin{equation}
H = \frac{1}{{\sqrt 2 }}\left( {\begin{array}{*{20}{c}}
1&1\\
1&{ - 1}
\end{array}} \right),{\kern 1pt} {\kern 1pt} T = \left( {\begin{array}{*{20}{c}}
1&0\\
0&{{e^{ - i{\pi  \mathord{\left/
 {\vphantom {\pi  4}} \right.
 \kern-\nulldelimiterspace} 4}}}}
\end{array}} \right),
\end{equation}
\begin{equation}
{\rm{CNO}}{{\rm{T}}_{12}} = \left( {\begin{array}{*{20}{c}}
1&0&0&0\\
0&1&0&0\\
0&0&0&1\\
0&0&1&0
\end{array}} \right).
\end{equation}

The experiments are carried out at room temperature on a Bruker AV-400 spectrometer (9.4 T).
The physical system is carbon-13 enriched chloroform (CHCL$_3$) dissolved in deuterated acetone. One ${}^1$H nucleus and one ${}^{13}$C nucleus of spin-1/2 are encoded as qubit $1$ and qubit $2$, respectively. The molecular structure and relevant parameters are shown in Fig. \ref{fig:molecule}. In the rotating frame, the internal Hamiltonian of the system can be written as
\begin{equation}
{\mathcal{H}_{int}} = \frac{\pi J}{2} \sigma_z^1\sigma_z^2,
\label{Hnmr}
\end{equation}
where $J=214.6\mathrm{Hz}$ is the scalar coupling strength between the two nuclei.

In NMR system, the thermal equilibrium state is a mixed state $\rho  = {{\left( {1 - \varepsilon } \right)} \mathord{\left/
 {\vphantom {{\left( {1 - \varepsilon } \right)} 4}} \right.
 \kern-\nulldelimiterspace} 4}\mathbb{I} + \varepsilon {\rho _\Delta }$, where $\mathbb{I}$ is the $4\times 4$ identity matrix, $\varepsilon\sim10^{-5}$ is the polarization, and $\rho_\Delta=4\sigma_z^1+\sigma_z^2+1/4\mathbb{I}$ is the deviation density matrix. The coefficients of $\sigma_z^1$ and $\sigma_z^2$ come from the fact that the gyromagnetic ratio of ${}^1$H is four times larger than ${}^{13}$C. Note that the dominant identity part is invariant under unital propagators, so we only consider the deviation part $\rho_\Delta$ in experiment.

\begin{table*}[htb]
\begin{tabular} {c||c|c|c|c|c}
  \hline
  Average Fidelity & $H_1$ & $H_2$ & $T_1$ & $T_2$ & $\rm{CNOT}_{12}$\\
  \hline
  \hline
  QPT: $\bar{F}(\Lambda, \mathcal{U})$      &$0.9903\pm 0.0005$  &$0.9850\pm0.0008$  & $0.9855\pm0.0007$ & $0.9937\pm0.0003$ &$0.9861\pm0.0006$\\
  \hline
  AUPT: $\bar{F}(\mathcal{V}, \mathcal{U})$ &$0.9826\pm 0.0010$  &$0.9863\pm0.0008$  & $0.9619\pm0.0023$& $0.9495\pm0.0018$ & $0.9350\pm0.0033$\\
  \hline
\end{tabular}
\caption{Average fidelities of the standard QPT and AUPT protocol compared to the theoretical gate, respectively. By randomly sampling 1000 input state in the 2-qubit pure state space for a given gate, we get one fidelity via Eq. (\ref{average_fidelity}). This procedure is repeated by 100 times, and the mean and standard deviation are used as the average fidelity and uncertainty in the table.}\label{fidelity_table}
\end{table*}

As reference, we firstly implement the standard QPT in experiment. The creation of all $16$ Pauli input states are realized by single-qubit rotations, free evolutions under the internal Hamiltonian, and $z$-gradient field pulses (to crush the unwanted non-zeroth coherence which is necessary in creating $II$) from the thermal equilibrium state. Then we apply the five gates to these Pauli input states via the following pulse sequences (pulses applied from right to left)
\begin{equation}
\begin{array}{l}
{H_1} = R_x^1\left( \pi  \right)R_y^1\left(
 \frac{\pi}{2} \right),\\
{H_2} = R_x^2\left( \pi  \right)R_y^2\left(
 \frac{\pi}{2} \right),\\
{T_1} = R_z^1\left(
 \frac{\pi}{4} \right),\\
{T_2} = R_z^2\left(
 \frac{\pi}{4} \right),\\
{\rm{CNOT_{12}}} = R_z^1\left(
 \frac{\pi}{2} \right) R_z^2\left(
 -\frac{\pi}{2} \right) R_x^2\left(
 \frac{\pi}{2} \right)U\left(
 \frac{1}{2J} \right) R_y^2\left(
 \frac{\pi}{2} \right).
\end{array}
\label{U pulse}
\end{equation}
The notation $R_{\hat{n}}^i\left( \theta \right)$ represents a single-qubit rotation on qubit $i$ along the $\hat{n}$-axis with the rotating angle $\theta$, and $U(t)$ represents the free evolution under the internal Hamiltonian in Eq. (\ref{Hnmr}) with time $t$.
The $z$-rotations in $T_1$ and $T_2$ can be decomposed by the formula ${R_z}\left( \theta  \right) = {R_x}\left( {{\pi  \mathord{\left/
 {\vphantom {\pi  2}} \right.
 \kern-\nulldelimiterspace} 2}} \right){R_y}\left( \theta  \right){R_x}\left( {{{ - \pi } \mathord{\left/
 {\vphantom {{ - \pi } 2}} \right.
 \kern-\nulldelimiterspace} 2}} \right)$. Note that $z$-rotations in NMR can also be realized more precisely by virtually varying the reference frame \cite{ryan2008liquid}, but this approach is not used in this experiment as it is better to apply imperfect pulses in order to address the stability of the AUPT protocol. Finally, by doing full QST, we reconstruct each gate in terms of an imperfect quantum channel $\Lambda$ as described in Eq. (\ref{StdQPT}).

For the AUPT protocol, starting from the thermal equilibrium state, we firstly create the pseudo-pure state (PPS)
\begin{align}\label{pps2}
\rho_{00}=\frac{1-\epsilon}{4}{\mathbb{I}}+\epsilon\ket{00}\bra{00}
\end{align}
using the spatial average technique \cite{cory1997ensemble,lu2011simulation}.
The other states $\left|01\right\rangle$, $\left|10\right\rangle$ and $\left|11\right\rangle$ are created from $\left|00\right\rangle$ by $\pi$ rotations. After applying one of the five gates, we measure the module of each element in $\mathcal{V}$ in Eq. (\ref{U0001_1}) by standard tomography of diagonal elements in NMR.
The relative phases within a given column correspond to the phases of single coherent terms, which is straightforward to read out in NMR as the spectrometer uses quadrature detection. In fact, all relative phases within one column for all five gates can be obtained in this way.

Next we need the relative phases between columns for each gate. We initialize $(\left|00\right\rangle+\sqrt{3}\left|01\right\rangle)/2$, $(\left|00\right\rangle+\sqrt{3}\left|10\right\rangle)/2$ and $(\left|10\right\rangle+\sqrt{3}\left|11\right\rangle)/2$ as the input states by applying $\pi/3$ rotations on the PPS state, which enables the reconstruction of all relative phases between columns for $H_1$, $H_2$, $T_1$ and $T_2$. However, $\mathrm{CNOT}_{12}$ is an exception. The application of $\mathrm{CNOT}_{12}$ to $\left|00\right\rangle+\left|10\right\rangle$ generates double quantum coherence $\left|00\right\rangle+\left|11\right\rangle$ which cannot be directly read out in NMR. The solution is to
apply another $\mathrm{CNOT}_{12}$ gate before detection to evolve double coherence back to single coherence, which may roughly double the error in $\mathrm{CNOT}_{12}$. Till now, we have successfully implemented the AUPT protocol for all five gates, and characterized each $\mathcal{V}$ in Eq. (\ref{UnitaryQPT}) individually.

\section{results and discussion}
\label{sec:results}

The reconstructions of each gate $H_1$, $H_2$, $T_1$, $T_2$ and $\rm{CNOT}_{12}$ via the standard QPT and AUPT protocols, as well as the theoretical results, are all shown in Fig. \ref{fig:results}. The five rows show the five gates, and the left, middle and right column are the theoretical, standard QPT, and AUPT results, respectively. Each subfigure shows the complete information of the target channel in the Pauli representation, as shown by the 16-by-16 matrix in Eq. (\ref{StdQPT}). Note that the AUPT results (the right column) are initially obtained via Eq. (\ref{U0001_1}) which is 4-by-4, and then converted to their equivalent 16-by-16 matrices in Eq. (\ref{StdQPT}) for fair comparisons with the other results. In each subfigure, the $y$-axis describes the input state in Eq. (\ref{StdQPT}), and $x$-axis describes the output state in the Pauli basis after applying the current channel. For example, the first column in each subfigure shows that when applying the channel to the input state $XX$, what the coefficients of the output state in Pauli basis are. From Fig. \ref{fig:results}, we see that the standard QPT results are closer to the theoretical predictions than the AUPT results.

To describe how closely that the practical channel $\Lambda$ approximates the theoretical channel $\mathcal{U}$ which is unitary in our case, one can use the value of diamond norm \cite{benenti2010computing} or average fidelity. Here we use the average fidelity between two channels, which is defined as
\begin{align} \label{average_fidelity}
\bar{F}(\Lambda, \mathcal{U}) = \int  \bra{\psi} \mathcal{U}^{\dagger} \Lambda (\ket{\psi} \bra{\psi}) \mathcal{U} \ket{\psi} d\mu(\psi),
\end{align}
where $d\mu(\psi)$ is the unitarily invariant distribution of pure states known as Fubini-Study measure \cite{emerson2005scalable}. For simplicity, we randomly sample 1000 $\ket{\psi}$'s from the 2-qubit pure state space, and replace the integral in Eq. (\ref{average_fidelity}) by the sum (with some normalization). The calculated average fidelities of the standard QPT protocol $\bar{F}(\Lambda, \mathcal{U})$ and AUPT protocol $\bar{F}(\mathcal{V}, \mathcal{U})$, both compared with the theoretical results $\mathcal{U}$, are shown in Table \ref{fidelity_table} for all the five gates. To get each average fidelity and its uncertainty, we randomly sample 1000 2-qubit pure states to get one value via Eq. (\ref{average_fidelity}) and repeat this procedure for 100 times. The average fidelity and uncertainty are defined as the mean and standard deviation of the 100 repetitions. The uncertainty for each gate is very small, which means 1000 samples are sufficient to estimate the average fidelity with a high precision.

Now let us discuss the error sources in two aspects. First of all, both of the standard QPT and AUPT results suffer the decoherence effect, imperfection of pulses, and state preparation and measurement (SPAM) errors. The decoherence is almost ignorable, as the gate implementation time is less than 3 ms, much shorter than the relaxation time scales which are at least 350 ms as shown in Fig. \ref{fig:molecule}. The imperfection of pulses such as over-rotation and under-rotation induce the SPAM errors, as well as the target gate infidelity. Just to clarify, it is hard for either the standard QPT or the AUPT protocol to distinguish the wanted gate error from the SPAM errors, but these two protocols both provide complete information of an unknown quantum channel. In contrast, the randomized benchmarking protocol \cite{emerson2007symmetrized} enables the separation of the gate error rate from the SPAM errors, but fruitless in fully characterizing the quantum channel.

Secondly, in Table \ref{fidelity_table} the AUPT results are worse than the standard QPT results (except $H_2$, for which we think the fluctuations in the SPAM error dominate the infidelity, and make it singular). The reason can be attributed to two factors. On one hand, the AUPT protocol is adaptive, that the next measurement relies on the previous one. It enables the propagation and amplification of the error to the latter experiments from the earlier experiments. On the other hand, to measure the relative phase via Eq. (\ref{phasemeasure}), we need to know the modules for each element and choose the single coherence  --- the only term that can be observed directly in NMR. An extreme case is the $\rm{CNOT}_{12}$ gate, that we have to apply it twice in order to evolve double coherence back to single coherence and observe the relative phase. That is why the AUPT result of $\rm{CNOT}_{12}$ is much worse than the case of standard QPT. Therefore, we conclude that AUPT indeed improves the efficiency significantly in characterizing an unknown quantum channel experimentally by assuming it is unitary, whereas it does have some drawbacks such as the two issues mentioned above.

\section{Conclusion}
\label{sec:conclusion}

In summary, we studied the quantum state tomography and unitary channel tomography via adaptive measurements. We showed that adaptive measurements can reduce the number of measurements when compared to non-adaptive measurements. In particular, we proved that pure state tomography can be accomplished using $2d-1$ measurements. By employing this idea, we demonstrated that $d^2+d-1$ measurements are sufficient to reconstruct a unitary process when the adaptive scheme is allowed.

Additionally, we implement our AUPT protocol for the universal gate set of quantum computing in a 2-qubit NMR system. Our results show that for local gates such as Hadamard and $T$ ($\pi/8$ phase) gates, high fidelities can be achieved using the AUPT protocol. For two-body gate such as the CNOT gate, the fidelity drops by some amount due to the accumulation of the errors in adaptively measuring the relative phases. Nevertheless, the AUPT protocol is still a useful tool in characterizing the unitary channels as it allows a significant reduction in terms of the required experiments, in particular for the local unitary channels.

\section{Acknowledgments}
\begin{acknowledgments}
We are grateful to the following funding sources: NSERC (N.Y., D.L., D.K., and B.Z.); CIFAR (B.Z.); National Natural Science Foundation of China under Grants No. 11175094 and No. 91221205 (K.L., and T.X.); National Basic Research Program of China under Grant No. 2015CB921002 (K.L., and T.X.). H.W., X.P., and J.D. would like to thank the following funding sources: NKBRP (2013CB921800 and 2014CB848700), the National Science Fund for Distinguished Young Scholars (11425523), NSFC (11375167, 11227901 and 91021005).
\end{acknowledgments}

\vskip 12pt

\end{document}